\documentclass[reprint,prl,twocolumn,superscriptaddress,showpacs,showkeys,floatfix,preprintnumbers]{revtex4-1}

\usepackage{braket}
\usepackage{ulem}
\usepackage{xargs}
\usepackage{diagbox}
\usepackage{mathtools}
\usepackage[english]{babel}
\usepackage[utf8]{inputenc}
\usepackage{amsmath,amssymb,braket,slashed,mathrsfs}
\usepackage{pifont}
\usepackage{MnSymbol}
\usepackage{graphicx}
\usepackage{float}
\usepackage{subfigure}
\graphicspath{{./fig/}}
\usepackage{tikz}
\usetikzlibrary{shapes,arrows,positioning}
\tikzset{decision/.style={diamond, draw, fill=blue!20, text width=4.5em, text badly centered, inner sep=0pt}}
\tikzset{block/.style={rectangle, draw, fill=blue!20, text width=10em, text centered, rounded corners, minimum width=3.5cm}}
\tikzset{block1/.style={rectangle, draw, fill=blue!20, text width=18.5em, text centered, rounded corners, minimum width=3.5cm}}
\tikzset{line/.style={draw, -latex, thick}}
\usepackage{color,hyperref}
\usepackage{multirow,array}

\usepackage{cleveref}

\newcommand{\pkuphy}{School of Physics, Peking University, Beijing 100871,
China}
\newcommand{\chep}{Center for High Energy Physics, Peking University, Beijing 100871, China}
\newcommand{\ccqm}{Collaborative Innovation Center of Quantum Matter, Beijing 100871, China}
\newcommand{\FRIB}{Facility for Rare Isotope Beams and Department of Physics and Astronomy,\\
Michigan State University, Michigan 48824, USA}
\newcommand{\gscaep}{Graduate School of China Academy of Engineering Physics, Beijing 100193, China}
\newcommand{\scnu}{Key Laboratory of Atomic and Subatomic Structure and Quantum Control (MOE),
 Institute of Quantum Matter, South China Normal University, Guangzhou 510006, China}
 \newcommand{\Gaziantep}{Faculty of Natural Sciences and Engineering, Gaziantep Islam Science and Technology University, Gaziantep 27010, Turkey}
 
 \newcommand{\SCNT}{Southern Center for Nuclear-Science Theory (SCNT), Institute of Modern Physics, Chinese Academy of Sciences, Guangdong 516000, China.}

\begin{document}
\title{Ab initio lattice calculation of nuclear magnetic dipole moments \\ with systematic error quantifications}

\author{Teng~Wang}\email{tenggeer@pku.edu.cn}\affiliation{\pkuphy}
\author{Serdar Elhatisari}\affiliation{\Gaziantep}
\author{Xu~Feng}\email{xu.feng@pku.edu.cn}\affiliation{\pkuphy}\affiliation{\chep}\affiliation{\ccqm}\affiliation{\SCNT}
\author{Dean Lee}\affiliation{\FRIB}
\author{Bing-Nan Lu}\email{bnlv@gscaep.ac.cn}\affiliation{\gscaep}
\author{Yuan-Zhuo Ma}\affiliation{\FRIB}\affiliation{\scnu}

%

\date{\today}

\begin{abstract}
Nuclear magnetic moments are sensitive probes of nuclear structure. However, their accurate quantitative description poses significant challenges, demanding both accurate nuclear and electromagnetic interactions as well as rigorous control of algorithmic uncertainties. 
Here, we present the first systematic calculation of magnetic dipole moments for selected light nuclei and aluminum isotopes within nuclear lattice effective field theory (NLEFT), an \textit{ab initio} framework applicable to medium-mass and heavy nuclei. Our calculations employ a lattice next-to-next-to-next-to-leading-order (N$^3$LO) chiral interaction together with electromagnetic currents consistently derived up to the two-body level. To achieve controlled predictions, we incorporate recently developed NLEFT algorithms and perform a comprehensive assessment of algorithmic uncertainties. Within the estimated uncertainties, our results are in good overall agreement with experiment and demonstrate that two-body currents are essential for reproducing the observed magnetic moments. We further benchmark our predictions against other \textit{ab initio} calculations for light nuclei ($A\leq12$).
Our work establishes a solid foundation for \textit{ab initio} studies of electroweak observables using methods that scale efficiently to medium-mass and heavy nuclei while demonstrating state-of-the-art accuracy.

\end{abstract}

\maketitle

\paragraph{Introduction}
Nuclear magnetic moments are precision probes of quantum many-body dynamics in atomic nuclei. While the magnetic moments of elementary particles provide some of the most stringent tests of fundamental theories such as quantum electrodynamics~\cite{Schwinger:1948}, nuclear magnetic moments arise from the interplay of single-particle motion and many-body correlations, encoding shell structure and valence configuration mixing in nuclei~\cite{Arima:1954,Blin-Stoyle:1956,Castel:1990,Ichikawa:2018,Vernon:2022}. 
Moreover, their strong sensitivity to the spin-isospin sector of nuclear interactions~\cite{Suzuki:2003,Munoz:2026} makes them particularly powerful constraints on underlying nuclear forces, complementing bulk observables such as binding energies and charge radii~\cite{Munoz:2026}.

Conventional descriptions of nuclear magnetic moments are based on mean-field and shell-model approaches employing phenomenological interactions~\cite{Arima:1954, Schmidt:1937,Castel:1990,Li:2018}. These methods successfully reproduce the overall systematics of magnetic moments~\cite{Schmidt:1937} and have elucidated important many-body mechanisms, including configuration mixing~\cite{Arima:1954} and meson-exchange currents~\cite{Miyazawa:1951,Villars:1952}. However, their reliance on phenomenological interactions and uncontrolled approximations introduces significant model dependence, making rigorous uncertainty quantification difficult and ultimately limiting their predictive power. This motivates \textit{ab initio} approaches based on fundamental interactions and systematically improvable many-body methods. At the most fundamental level, lattice quantum chromodynamics (QCD) provides a direct solution of QCD with systematically controllable uncertainties, including statistical and discretization errors. Magnetic moments have been computed from first-principles QCD for light nuclei with $A\le3$~\cite{Beane:2014,Chang:2015}, including the deuteron, $^{3}$H, and $^{3}$He~\cite{Beane:2014}. However, direct lattice QCD calculations remain limited to very light nuclei because of the exponential computational complexity.



Nuclear \textit{ab initio} methods rooted in chiral effective field theory ($\chi$EFT) offer an efficient alternative to direct lattice QCD, systematically linking low-energy interactions to QCD symmetries through a controlled power counting~\cite{Weinberg:1978, Hergert:2020}. 
Over the past decades, remarkable progress has greatly extended the reach of \textit{ab initio} nuclear theory. Systematic magnetic moment calculations have been carried out for light nuclei using Quantum Monte Carlo methods~\cite{ChambersWallPRL,ChambersWallPRC,Lonardoni:2018} and the no-core shell model~\cite{Forssen:2009,NCSM_B,NCSM_C}, and for medium-mass and heavy nuclei using the valence-space in-medium similarity renormalization group~\cite{Vernon:2022,VS-IMSRG1,VS-IMSRG2,VS-IMSRG3,Miyagi:2024} and coupled-cluster theory~\cite{Acharya:2024}. While these approaches successfully reproduce magnetic moments for several benchmark nuclei, noticeable discrepancies remain, underscoring the need for systematic uncertainty quantification.

As an advanced \textit{ab initio} approach, nuclear lattice effective field theory (NLEFT) combines $\chi$EFT with lattice Monte Carlo techniques to solve nuclear many-body problems~\cite{NLEFT_review_Dean,NLEFT_review_Ulf}. Treating all nucleons as dynamical degrees of freedom with only mild power-law scaling of the computational cost and employing imaginary-time projection, NLEFT provides a unified framework for describing nuclei from the light- to the heavy-mass region. NLEFT has achieved systematic descriptions of nuclear binding energies, spectra, and bulk properties across a wide mass range~\cite{Lahde:2013,essential_elements,Niu:2025,WFM,Hildenbrand:2025,Song:2025,Beisotope,C12,Ren:2025,Konig:2023,Brinson:2026}. However, observables that probe detailed many-body structures, particularly electromagnetic (EM) observables, remain largely unexplored.
The primary obstacle is the severe Monte Carlo sign problem associated with realistic chiral interactions. In practical calculations, this is mitigated by introducing an auxiliary sign-problem-free Hamiltonian and treating the difference perturbatively~\cite{PTmethod1,LiuJun}. While this strategy is highly successful for energies, it becomes significantly less efficient for EM observables, whose accurate calculation is hindered by enhanced excited-state contamination and statistical noises~\cite{Wang:2026}. Another obstacle is the rotational symmetry breaking induced by the discrete cubic group~\cite{Johnson:1982,Berg:1983,Mandula:1982} and the associated enhanced lattice artifact for spin-dependent observables such as the magnetic moment~\cite{NLEFT_rotationsymmetry2}. To achieve an accurate lattice calculation, the reduction and quantification of such artifact is required.

To overcome this long-standing limitation, we recently developed a multi-reference trial-state approach that efficiently samples shell-model configurations within NLEFT~\cite{Wang:2026}. By dramatically accelerating imaginary-time convergence, the method suppresses both excited-state contamination and the statistical uncertainties associated with perturbative corrections, enabling systematic \textit{ab initio} calculations of EM observables. We show that, by properly optimizing the trial wave functions and restoring rotational symmetry, \textit{ab initio} lattice calculations can achieve an accuracy for spin-dependent observables comparable to that of continuum-based methods that possess exact rotational symmetry.

Here we present the first systematic NLEFT calculation of magnetic dipole moments for selected nuclei up to $A=12$. As stringent probes of nuclear wave functions, magnetic moments provide an ideal benchmark for assessing the accuracy of many-body methods. They also receive important contributions from two-body EM currents~\cite{2Bexpression2,Schiavilla:2019,Pastore:2013,ChambersWallPRC,ChambersWallPRL,Lonardoni:2018,Miyagi:2024}, making their accurate description an essential step toward reliable predictions of electroweak observables such as neutrinoless double-$\beta$ decays~\cite{Menendez:2011,Engel:2014,Engel:2016,Dolinski:2019}. We further extend the study to the aluminum isotopic chain,  which is of current experimental
interest at radioactive-beam facilities~\cite{Brinson:2026} and meanwhile illustrating the favorable mass scaling of NLEFT. Particular emphasis is placed on quantifying lattice uncertainties to enable precision predictions.
 


\noindent
\paragraph{Methodology}\label{section:method}

The master formula for the lattice calculation of  the magnetic dipole moment is  
\begin{equation}
\label{eq:Master}
   \mu_{\mathrm{latt}} = \lim_{L_t \rightarrow \infty}\frac{\langle \Psi^T_{J,M=J}|\mathcal{M}^{L_t/2}\mu_z \mathcal{M}^{L_t/2}|\Psi^T_{J,M=J}\rangle}{\langle \Psi^T_{J,M=J}|\mathcal{M}^{L_t}|\Psi^T_{J,M=J}\rangle}.
\end{equation}
In the above equation, $\mu_z$ is the third component of the magnetic dipole moment operator $\boldsymbol{\mu} = (\mu_x, \mu_y, \mu_z)$. 
$|\Psi^T_{J,M=J}\rangle $ is the trial state carrying the total angular moment $J$ and the magnetic quantum number $M=J$ as the lattice spacing $a\rightarrow 0$ and box size $L\rightarrow \infty$. $\mathcal{M}=:e^{-H a_t}:$ is the transfer matrix, with $H$ the nuclear many-body Hamiltonian, $a_t$ the temporal lattice spacing, and $:\ :$ the normal-order operator. $L_t$ is the number of transfer matrices. Since all nuclei studied in this work are either the ground state or the lowest-energy state of a given spin $J$, the above single-channel formula is sufficient to remove excited-state contamination as $L_t$ is large enough. 

The magnetic moment operator $\boldsymbol{\mu}$ can be expressed as the curl of the EM current  $\boldsymbol{j}(\boldsymbol{k})$ in the limit of vanishing momentum transfer $\boldsymbol{k}$,
\begin{equation}
\label{eq:Definition}
    \boldsymbol{\mu} = \frac{1}{2 i}\mathop{\lim}_{k\rightarrow 0}\nabla_{\boldsymbol{k}}\times \boldsymbol{j}(\boldsymbol{k}).
\end{equation}
 In $\chi$EFT, one- and two-body EM currents have been constructed up to N$^3$LO~\cite{EMcurrent_Pastore, EMcurrent_Krebs1, EMcurrent_Krebs2, EMcurrent_Krebs3}, and the latter gives two-body corrections to the magnetic moment operator \cite{2Bexpression1, 2Bexpression2, EMcurrent_Pastore}. In this work, we consider the leading one- and two-body terms,
 \begin{equation}
     \boldsymbol{\mu} = \boldsymbol{\mu}_{1\mathrm{N}}+\boldsymbol{\mu}_{2\mathrm{N}}.
 \end{equation} $\boldsymbol{\mu}_{1\mathrm{N}}$ is the one-body dipole moment operator whose form is well-known,
\begin{equation} 
\label{eq:Mag1B}
\boldsymbol{\mu}_{1\mathrm{N}}=\mu_N\sum_{n}\left(\frac{g_S+g_V\tau_n^3}{2}\boldsymbol{\sigma}_{n} +\frac{1+\tau_n^3}{2}\boldsymbol{l}_{n}\right).
\end{equation}
In the above,  $\mu_N=e/2m_N$ is
the nucleon magneton. $\boldsymbol{\sigma}_{n}$ and $\boldsymbol{l}_{n}$ are the spin and the orbital angular momentum of the $n$th nucleon.  $g_S= 0.880$  and $g_V = 4.706 $ are the isoscalar and isovector $g$ factors, respectively~\cite{gfactor}. For the two-body part $\boldsymbol{\mu}_{2\mathrm{N}}$, we employ its leading power-counting term in this work and leave its explicit expression in the Supplementary Material. For numerical calculations, both $\boldsymbol{\mu}_{1\mathrm{N}}$ and $\boldsymbol{\mu}_{\mathrm{2N}}$ are realized through a direct lattice regularization.

In this work, we use the recently developed high-fidelity next-to-next-to-next-to-leading-order (N$^3$LO) chiral Hamiltonian $H=H_\chi$, with lattice spacing $a=1.32$~fm~\cite{WFM}. Calculations are  performed at box sizes $L=9a$ to $L=11a$ to balance computational costs and finite-volume effects.  Due to the severe Monte Carlo sign problem, a fully non-perturbative  calculation with $H_\chi$ is extremely hard. To mitigate this difficulty, we introduce a simple leading-order (LO)  Hamiltonian $H_S$ (see Ref.~\cite{WFM}), which can be simulated non-perturbatively, and treat the difference $H_\chi-H_S$ as a perturbation. Observables of interest are then calculated order by order perturbatively~\cite{PTmethod1,LiuJun}. For the calculation of the magnetic dipole moment, we retain the contribution of the one-body term $\boldsymbol{\mu}_{1\mathrm{N}}$ up to the first order, and neglect  perturbative  corrections beyond the leading order for  $\boldsymbol{\mu}_{2\mathrm{N}}$, due to its relatively small magnitude, i.e.,
\begin{equation}
\label{eq:PTmethod}
    \mu_{\mathrm{latt}} \approx \mu_{\mathrm{1N}}^{(0)}+\mu_{\mathrm{1N}}^{(1)}+\mu_{\mathrm{2N}}^{(0)},
\end{equation}
where the superscript denotes the perturbative order. Higher-order perturbative terms neglected here are computationally demanding and assumed to be small. Their impact on the final result will be investigated in future works.

\paragraph{Uncertainty quantification}
In this work, we consider two sources of uncertainties: the statistical fluctuation and the associated extrapolation error against the projection time $\tau=L_t a_t$, as well as the rotational symmetry breaking effect inherent in lattice calculations. .

For the first source of uncertainty, Ref.~\cite{Wang:2026} points out the potentially large excited-state contamination and extrapolation errors versus $\tau$ for the magnetic dipole moment, due to its sensitivity to nuclear valence configurations.  To suppress excited-state effects, we  use the following multi-reference shell-model wave function proposed in Ref.~\cite{Wang:2026},
\begin{equation}
\label{eq:TrialState}
  |\Psi^T_{J, M=J}\rangle=\sum_{n=1}^{n_{\mathrm{max}}} a_n |\Psi_{M=J}^{n}\rangle. 
\end{equation}
In Eq.~(\ref{eq:TrialState}), $|\Psi_{M=J}^n\rangle$ with  $n=1,\cdots, n_{\mathrm{max}}$ are  single Slater determinants belonging to the  same shell-model ground-state configuration with magnetic quantum number $M=J$. The coefficients $a_{n}$ are properly chosen to ensure that $|\Psi^T_{J,M=J}\rangle$ has the definite spin $J$ in the continuum limit. Compared to a single Slater determinant which generally mixes nuclear spins different from $J$, our trial state significantly reduces the excited-state contamination   for the calculation of the magnetic moment, allowing a reliable imaginary-time extrapolation with controlled uncertainties. In the Supplementary Material, we provide a more comprehensive discussion on the trial-state construction, the mechanism of excited-state suppression and the improvement of extrapolation uncertainty.

For lattice calculations, the finite box length $L$ and lattice spacing $a$ break the rotational
symmetry, reducing the full rotational group to the finite group of cubic rotations~\cite{Johnson:1982,Berg:1983,Mandula:1982}. Therefore,   lattice calculation of observables always suffers from the mixing of different angular momenta, and such unphysical artifact is enhanced for for spin-dependent observables such as the magnetic moment~\cite{NLEFT_rotationsymmetry2}. To quantify the influence, we vary the size $L$ of the box to see how $\mu$ depends on $L$. For the lattice spacing effect, since it is nontrivial to vary $a$ and take it to zero in nonrenormalizable field theories such as $\chi$EFT~\cite{NLEFT_rotationsymmetry1}, here we adopt an indirect method of Ref.~\cite{NLEFT_rotationsymmetry2} to access the magnitude of the associated rotational symmetry breaking. We find that for light nuclei of $sp$-shell, the impact of rotational symmetry breaking on the magnetic moment is negligible for $a=1.32$fm and $L\ge9a$. For the heavier aluminum  isotopes, the impact is enhanced but still controlled at percent level, and we assign a $2\%$ total uncertainty to account for that.   The details of the analysis are also left in the Supplementary Material.

\paragraph{Results}

\begin{figure}[htb]
\centering
\includegraphics[width=0.46\textwidth,angle=0]{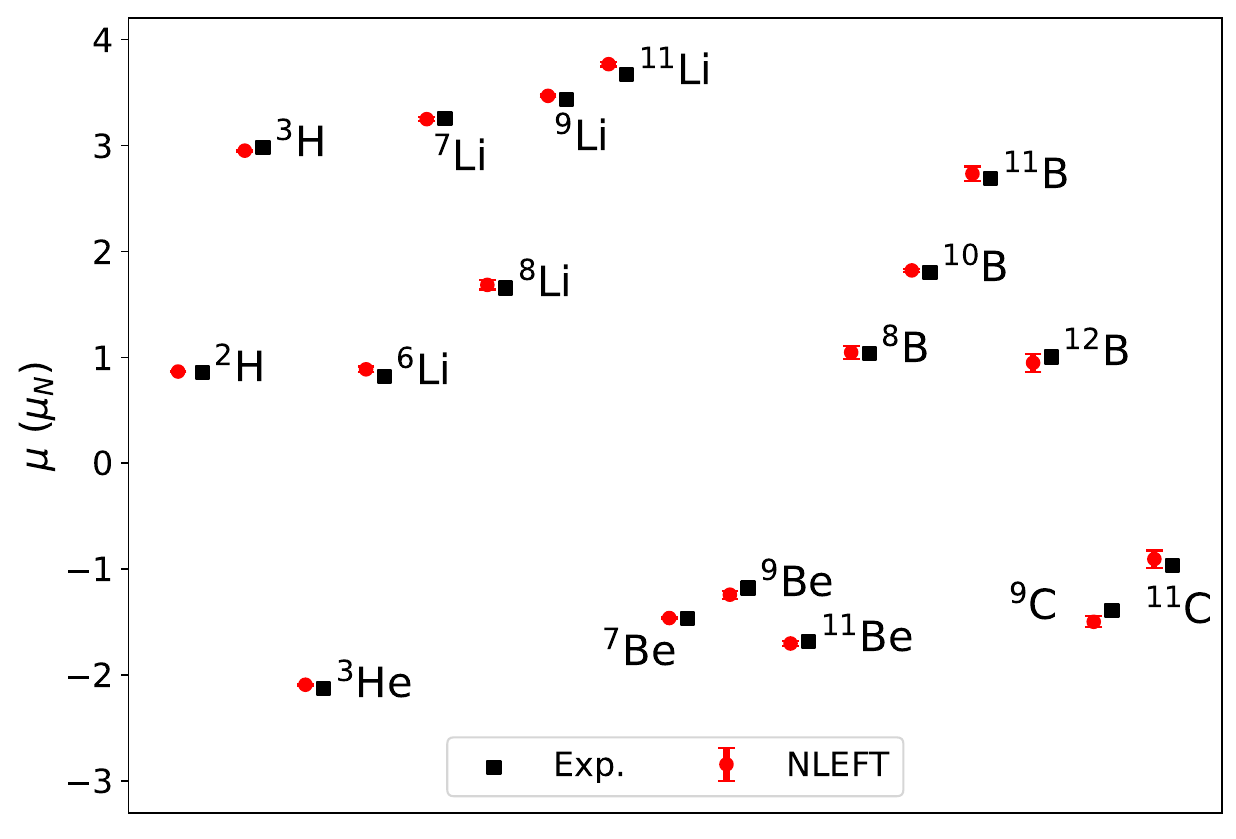}
\caption{Global comparison between the full NLEFT results and experimental results, denoted by the red and black points respectively, for the ground state magnetic moments of selected $A\le 12$ nuclei. The error bar denotes theoretical uncertainty.}
\label{fig:Mag_vs_exp_spshell}
\end{figure}

\begin{figure}[htb]
\centering
\includegraphics[width=0.48\textwidth,angle=0]{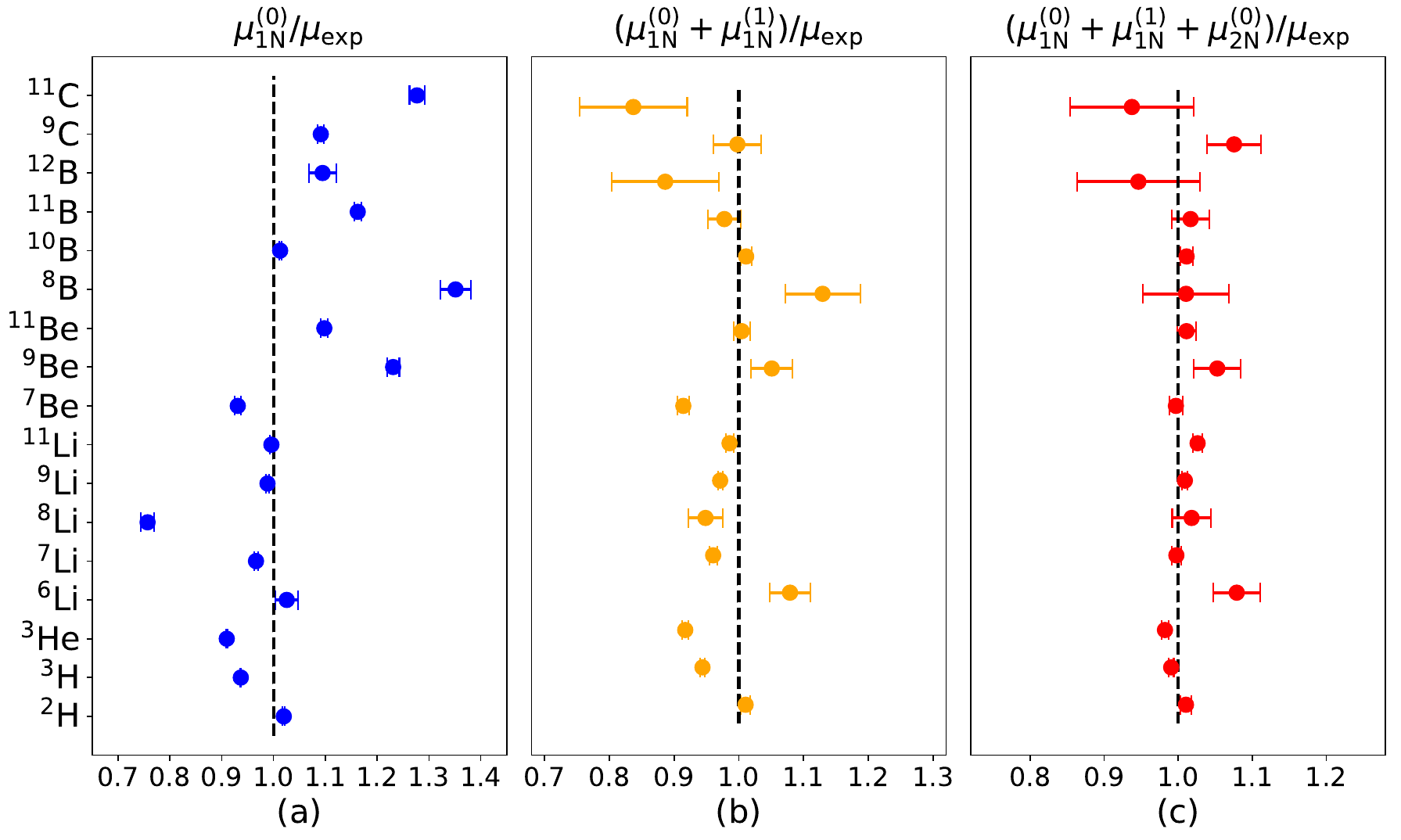}
\caption{The ratio between the magnetic moment calculated on the lattice and the experiment. Panels (a), (b) and (c) correspond to $\mu_{\mathrm{1N}}^{(0)},\mu_{\mathrm{1N}}^{(0)}+\mu_{\mathrm{1N}}^{(1)}$ and $\mu_{\mathrm{latt}} = \mu_{\mathrm{1N}}^{(0)}+\mu_{\mathrm{1N}}^{(1)}+\mu_{\mathrm{2N}}^{(0)}$, respectively. The error bar denotes theoretical uncertainty.}
\label{fig:Mag_3terms}
\end{figure}

\begin{figure*}
\centering
\includegraphics[width=0.8\textwidth,angle=0]{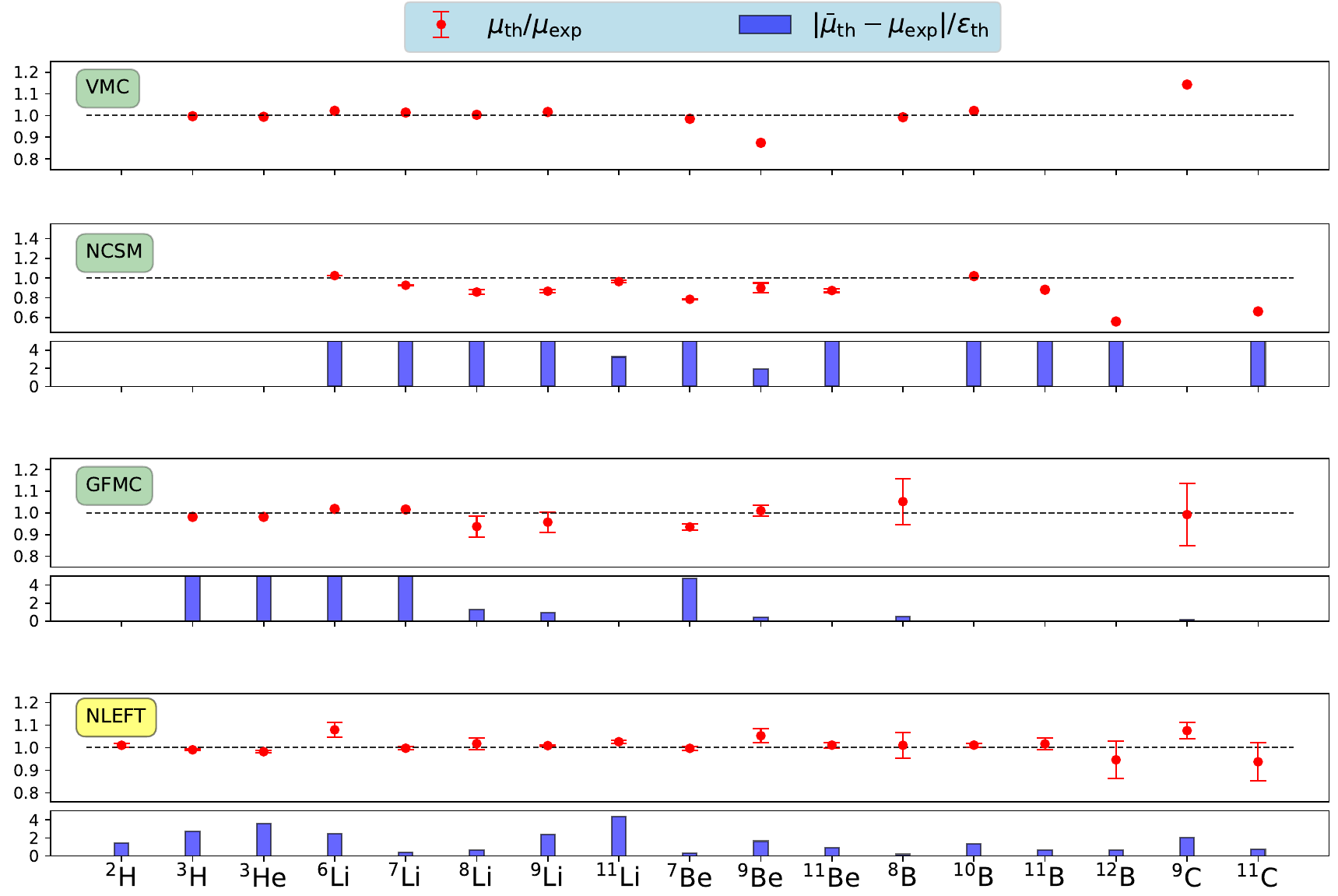}
\caption{The ratio between the theoretical prediction and the experimental value from different $ab$ $initio$ methods, for the ground state magnetic moment of selected $A\le 12$ nuclei. The four subpolts from top to bottom  represent NLEFT, VMC, GFMC and NCSM methods, respectively. The red point represents the central value of the ratio and the error bar denotes theoretical uncertainty. The blue bar in the lower part of each subplot represents relative deviation from the experiment, as explained in the main text.  For VMC and GFMC, we adopt the result of Ref~\cite{ChambersWallPRL, ChambersWallPRC} using the NV2+3-Ia* interaction~\cite{NV2+3}. For NCSM, we employ the result of Ref.~\cite{Forssen:2009,NCSM_B,NCSM_C} based on the INOY interaction~\cite{INOY}. }
\label{fig:Methods_compare_spshell}
\end{figure*}

We first present our result for the ground state magnetic moment of selected $A\le 12$ nuclei. Figure~\ref{fig:Mag_vs_exp_spshell} shows the global comparison between the full NLEFT result and  the experiment~\cite{Purcell:2010,Tilley:2002,Tilley:2004,Borremans:2005,Neugart:2008,Williams:1970, Millman:1939,Wolber:1970},  and we observe an overall  good agreement for these  $sp$-shell light nuclei.  To see the origin of this agreement and to gain the insight into the interplay between nuclear forces and currents,  we calculate the ratios $\mu^{(0)}_{1\mathrm{N}}/\mu_{\mathrm{exp}}$, $(\mu^{(0)}_{1\mathrm{N}}+\mu^{(1)}_{1\mathrm{N}})/\mu_{\mathrm{exp}}$ and $(\mu^{(0)}_{1\mathrm{N}}+\mu^{(1)}_{1\mathrm{N}}+\mu^{(0)}_{\mathrm{2N}})/\mu_{\mathrm{exp}}$ in order for all the light nuclei, and we present the result in Figure~\ref{fig:Mag_3terms}. For $\mu_{1\mathrm{N}}^{(0)}$ calculated from the simple Hamiltonian $H_S$, due to the miss of essential spin-isospin correlations, the ratio is widely scattered and deviates significantly from unity for most nuclei. Incorporating higher-order interactions yields systematic improvement as can be seen in panel (b). The correction is especially important for $^8$Li, $^9$Be $^8$B and $^{11}$C, contributing $\sim$30\% of the full result. In panel (c), remaining discrepancies (e.g., for $^3$H, $^3$He, and $^7$Be) are largely resolved by two-body currents. Although the
two-body current contributes mostly up to $\sim$10$\%$ level of the total moment, it is precisely this correction that performs the final fine-tuning bringing the
lattice results into agreement with experiment for these
nuclei; this is, to our knowledge, the first time that two-body EM currents have been included in an NLEFT calculation, and our results demonstrate that they are an
indispensable ingredient for the quantitative description
of nuclear magnetic moments, consistent with previous $ab\ initio$ calculations~\cite{2Bexpression2,Schiavilla:2019,Pastore:2013,ChambersWallPRC,ChambersWallPRL,Lonardoni:2018,Miyagi:2024}. Note that the result in some cases such as $^{11}$Li  still differs from the experiment by about $3\sigma$, which can be hopefully mitigated by either including higher-order chiral interactions and currents, or by analyzing remaining theoretical uncertainties missed in current work.

In Figure~\ref{fig:Methods_compare_spshell}, we compare the above NLEFT result with  predictions made by  three different $ab\ initio$ methods, including Variational Monte Carlo (VMC), Green's function Monte Carlo (GFMC) and no-core shell model (NCSM). For clarity, we again plot the theory-to-experiment ratios for all methods separately, represented by the red points in the four subplots in Figure~\ref{fig:Methods_compare_spshell}. For each method, we  focus on both the central value $\bar{\mu}_{\mathrm{th}}$ and the theoretical uncertainty $\epsilon_{\mathrm{th}}$, and we plot the relative deviation from the experiment, i.e. $|\bar{\mu}_{\mathrm{th}}-\mu_{\mathrm{exp}}|/\epsilon_{\mathrm{th}}$, as the blue bar in the lower part of each subplot (regions beyond 5$\sigma$ are trimmed for drawing).  Among the different methods, VMC gives no estimation of uncertainty, while the uncertainty of NCSM is very small. Since the absolute deviation $|\bar{\mu}_{\mathrm{th}}-\mu_{\mathrm{exp}}|$ of certain nuclei exceeds $10\%$ for both methods, the above mentioned missing and underestimation of uncertainty undermines their agreement with the experiment. For GFMC, we observe an overall smaller absolute discrepancy, but the uncertainty of selected nuclei, such as $^{3}$H, $^{3}$He, $^{6}$Li and $^7$Li, is also missed.  In comparison, the NLEFT results exhibit globally improved agreement with experiment, owing to well-controlled absolute deviations and moderate relative uncertainties (generally smaller than 3$\sigma$; the largest deviation for $^{11}$Li remains smaller than 5$\sigma$). We emphasize that, among all the methods compared here, NLEFT is the only one that provides a  reliable estimate of the theoretical uncertainty
for every nucleus considered. This capability of rigorous
uncertainty quantification, rooted in the systematically
improvable nature of $\chi$EFT and the statistically controlled lattice Monte Carlo, is a decisive advantage for
turning magnetic moments into quantitative constraints
on nuclear forces and currents.

\begin{figure}[htb]
\centering
\includegraphics[width=0.48\textwidth,angle=0]{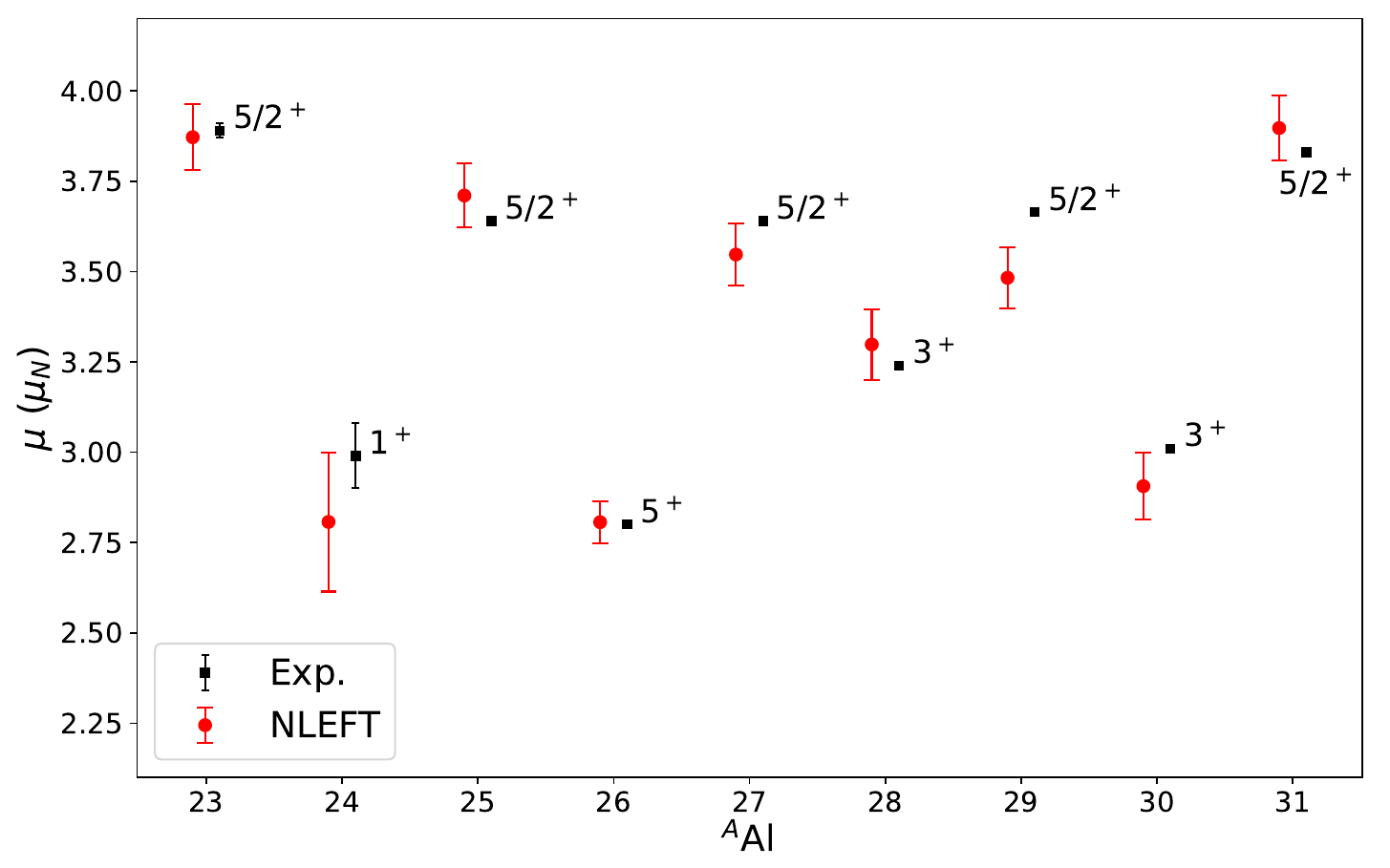}
\caption{The magnetic moments of the aluminum isotopes versus the mass number $A$. The red point represents the full result of NLEFT and the red bar denotes the theoretical uncertainty. The black point represents the experimental value and the black bar denotes the experimental error ~\cite{Ozawa:2006,Minamisono:1976,Cooper:1996,Nishimura:2007,Al33}.}
\label{fig:Mag_Al}
\end{figure}

Experimental interest in aluminum isotopes has increased in recent years~\cite{Brinson:2026}, including $^{22,23}$Al as the candidate of proton-halo nucleus~\cite{Al22, Al23},   and $^{33,34}$Al at the border of the island of inversion~\cite{Al33,Al34_1,Al34_2}.  Theoretically, the structure of middle shell nuclei like aluminum is sensitive to nuclear correlations, which is suitable for testing the nuclear many-body method~\cite{VS-IMSRG2}.  In Figure~\ref{fig:Mag_Al}, we present our prediction of magnetic moments along aluminum isotopes with $23\le A\le 31$. Compared to the experiment, the lattice result not only reproduces the evolutionary trend against the mass number, but also agrees with  experiment within $2\sigma$ for all isotopes. Note that for $^{24}$Al, since its ground-state magnetic moment is currently unknown experimentally, we calculate its $1^+$ excited state  instead and the statistical error is significantly larger.

\noindent
\paragraph{Summary}
In this work, we have presented a systematic NLEFT calculation of nuclear magnetic dipole moments for selected light nuclei and aluminum isotopes. 
In
the light sector, our results underscore the essential role
of many-body correlations and two-body currents. They
also demonstrate improved agreement with experiment
compared to other \textit{ab initio} methods, thanks to careful
uncertainty quantification. The good agreement for $sd$-shell aluminum isotopes also highlights the promise of
NLEFT for medium-mass nuclei. Moreover, the lattice
implementation of two-body electromagnetic currents established here, together with their systematic validation against precise magnetic-moment data, represents
a step toward NLEFT calculations of the broader class
of current-dependent electroweak observables.

Studies of nuclear shell structure and its evolution far from stability have deeply challenged the shell-model paradigm of nuclear physics and refined our understanding of nuclear  forces~\cite{Otsuka:2020}.  The magnetic dipole moment of key isotopic chains with protons forming stable configurations near magic number, including calcium~\cite{Garcia:2015}, copper~\cite{Ichikawa:2018} and indium~\cite{Vernon:2022}, is an important probe for visualizing and identifying the shell evolution. A systematic survey of such chains across the medium- and heavy-mass region would help interpret the abrupt changes and unexpected evolution of magnetic moments observed experimentally~\cite{Vernon:2022}, and would form a stringent test of NLEFT in heavier exotic systems. Reaching this region, however, requires overcoming several challenges that lie beyond the scope of the present work, including the rapid growth of computational cost with mass number, the Monte Carlo sign problem, and the fine-tuning of three-nucleon forces. Nevertheless, NLEFT has already achieved promising results for other observables up to medium-mass and heavy nuclei, such as binding energies and charge radii~\cite{WFM, Niu:2025, Hildenbrand:2025}, which gives us confidence that the precise calculation of magnetic moments in this region is a challenging but highly promising avenue for future exploration.

A key strength distinguishing the $ab\ initio$ method from phenomenological nuclear models is its ability to quantify theoretical uncertainties. In this study, we focus on the uncertainty induced by the lattice method itself, i.e., we construct trial states with desired nuclear spin to control the imaginary-time extrapolation uncertainty, and  analyze the rotational symmetry breaking effect and its impact on the magnetic moment. These efforts lay the foundation for systematic uncertainty quantification of nuclear electromagnetic properties in the framework of NLEFT. With the enhancement of computing power in the future, calculations can be performed on larger volumes with finer lattice spacings and larger statistics,  which would reduce the impact of the above effects and necessitates the  propagation of full uncertainties. Future work is needed to include the remaining errors missed here, including those induced by the perturbation method and $\chi$EFT truncations.


\begin{acknowledgments}
{\bf Acknowledgments}

X.F. and T.W. were supported in part by NSFC of China under Grants No. 12125501 and No. 12550007.
B.N.L. was supported by the Science Challenge Project (No. TZ2025012), NSAF No. U2330401 and National Natural Science Foundation of China with Grant
Nos. 12275259, 12547105.
S.E. was supported in part by Scientific and Technological Research Council
of Turkey (TUBITAK project no. 123F464). D.L. and Y.-Z.M. acknowledge support from the U.S. Department of Energy grants DE-SC0013365, DE-SC0023175, and DE-SC0026198.

\end{acknowledgments}

\bibliography{ref}

\clearpage

\setcounter{page}{1}
\renewcommand{\thepage}{Supplementary Information -- S\arabic{page}}
\setcounter{table}{0}
\renewcommand{\thetable}{S\,\arabic{table}}
\setcounter{equation}{0}
\renewcommand{\theequation}{S\,\arabic{equation}}
\setcounter{figure}{0}
\renewcommand{\thefigure}{S\,\arabic{figure}}

\begin{widetext}
\section{Supplementary Material}

\subsection{A. Lattice realization of the magnetic dipole moment operator}
\label{sectionA}

To define the magnetic dipole operator on the lattice, we first introduce the nucleon creation operator $a_{i,j}^\dagger(\boldsymbol{n})$ and the annihilation operator $a_{i,j}(\boldsymbol{n})$, where $\boldsymbol{n}=(n_x, n_y, n_x)$ is the lattice coordinate and $i,j=0,1$ are spin and isospin indices, respectively. We follow the convention that $i=0, 1$ denotes spin up and spin down while $j=0, 1$ denotes proton and neutron. We then introduce the following density operators,
\begin{equation}
\begin{aligned}
    &\rho(\boldsymbol{n}) = \sum_{i,j} a^\dagger_{i,j}(\boldsymbol{n})a_{i,j}(\boldsymbol{n}),\\
    &\rho_{S,0}(\boldsymbol{n}) = \sum_{i,i'}\sum_{j}a^\dagger_{i,j}(\boldsymbol{n})[\sigma_S]_{ii'}a_{i',j}(\boldsymbol{n}),\\
    &\rho_{0,I}(\boldsymbol{n}) = \sum_{i}\sum_{j,j'}a^\dagger_{i,j}(\boldsymbol{n})[\tau_I]_{jj'}a_{i,j'}(\boldsymbol{n}),\\
    &\rho_{S,I}(\boldsymbol{n}) = \sum_{i,i'}\sum_{j,j'}a^\dagger_{i,j}(\boldsymbol{n})[\sigma_S]_{ii'}[\tau_I]_{jj'}a_{i',j'}(\boldsymbol{n}),
\end{aligned} 
\end{equation}
as well as the  current operator for protons,
\begin{equation}
\begin{aligned}
    \theta_{k}(\boldsymbol{n}) = -\frac{i}{2}\sum_{i}(a^\dagger_{i,0}(\boldsymbol{n})\partial_{k}a_{i,0}(\boldsymbol{n})-\partial_{k}a^\dagger_{i,0}(\boldsymbol{n})a_{i,0}(\boldsymbol{n})).
\end{aligned}
\end{equation}
The Pauli matrices $\sigma_S$ and $\tau_I$ act on the spin and isospin space respectively, with $S,I\in\{x,y,z\}$. The partial derivative $\partial_k$ is defined through the discretized Fourier transformation to  suppress lattice  artifacts,
\begin{equation}
\label{eq:derivative_lat}
        \partial_k a_{i,j}(\boldsymbol{n}) =i\sum_{\boldsymbol{p}}p_k e^{i\boldsymbol{p}\cdot \boldsymbol{n}}\tilde{a}_{i,j}(\boldsymbol{p}),
\end{equation}
with $\tilde{a}_{i,j}(\boldsymbol{p})$ the nucleon annihilation operator in the momentum space. The dimensionless lattice momentum $\boldsymbol{p} = 2\pi a \boldsymbol{n}_p/L$ is summed over the first Brillouin zone.

 Using the notations above, the $k$th component of the one-body dipole moment operator Eq.(\ref{eq:Mag1B}) on the lattice is 
 \begin{equation}
 \label{eq:Mag1B_lat}
     \mu_{ 1\mathrm{N}, k} = \mu_N\sum_{\boldsymbol{n}}\left(\frac{g_S \rho_{k,0}(\boldsymbol{n})+g_V\rho_{k,z}(\boldsymbol{n})}{2}+\epsilon_{ijz} n_i \theta_j(\boldsymbol{n})\right),
 \end{equation}
with $\epsilon_{ijk}$ the Levi-Civita symbol. The first term in Eq.~(\ref{eq:Mag1B_lat}) corresponds to the nucleon spin term in  Eq.~(\ref{eq:Mag1B}), while the second term corresponds to the orbital motion of protons. The repeated indices $i,j$ are implicitly summed. 

For the  two-body dipole moment operators $\boldsymbol{\mu}_{\mathrm{2N}}$, it consists of the intrinsic term and the Sachs term, 
\begin{equation}
\label{eq:mag_decomp}
\boldsymbol{\mu}_{\mathrm{Intr}} =\sum_{i<j}^A\boldsymbol{\mu}^{ij}_{\mathrm{intr}},\quad \boldsymbol{\mu}_{\mathrm{Sachs}} =\sum_{i<j}^A\boldsymbol{\mu}^{ij}_{\mathrm{Sachs}},
\end{equation}
 their coordinate-space expressions are as follows in the continuum,
\begin{equation} 
\label{eq:mag_continuum}
\begin{aligned}   &\boldsymbol{\mu}^{ij}_{\mathrm{intr}}(\boldsymbol{r}_i-\boldsymbol{r}_j) = -\frac{e g^2_A m_\pi}{32\pi F_\pi^2}(\boldsymbol{\tau}_i\times\boldsymbol{\tau}_j)_z\left[\left(1+\frac{1}{m_\pi r_{ij}}\right)[(\boldsymbol{\sigma}_i\times\boldsymbol{\sigma}_j)\cdot\hat{\boldsymbol{r}}_{ij}]\hat{\boldsymbol{r}}_{ij}-(\boldsymbol{\sigma}_i\times\boldsymbol{\sigma}_j)\right]e^{-m_\pi r_{ij}}\\
&\boldsymbol{\mu}^{ij}_{\mathrm{Sachs}}(\boldsymbol{r}_i-\boldsymbol{r}_j) = -\frac{e g_A^2}{32\pi F_\pi^2}(\boldsymbol{\tau}_i\times\boldsymbol{\tau}_j)_z\left[[S_{ij}(\hat{\boldsymbol{r}}_{ij})h(r_{ij})+(\boldsymbol{\sigma}_i\cdot\boldsymbol{\sigma}_j)]\frac{m_\pi^2 e^{-m_\pi r_{ij}}}{3r_{ij}}-\frac{4\pi}{3}\boldsymbol{\sigma}_i\cdot\boldsymbol{\sigma}_j\delta(\boldsymbol{r}_{ij})\right](\boldsymbol{R}_{ij}\times\boldsymbol{r}_{ij}).
\end{aligned}
\end{equation}
In the expressions above, $g_A=1.287$ is the axial-vector coupling constant, $F_\pi=92.2$ MeV is the pion decay constant and $m_\pi=134.98$ MeV is the pion mass. We  defined $\boldsymbol{r}_{ij}=\boldsymbol{r}_{i}-\boldsymbol{r}_{j}$ and $\boldsymbol{R}_{ij} =(\boldsymbol{r}_i+\boldsymbol{r}_j)/2$. $\hat{\boldsymbol{r}}_{ij}$ is the unit vector of $\boldsymbol{r}_{ij}$. We also definecd $S_{ij}(\hat{\boldsymbol{r}}_{ij})=3(\hat{\boldsymbol{r}}_{ij}\cdot\boldsymbol{\sigma}_i)(\hat{\boldsymbol{r}}_{ij}\cdot\boldsymbol{\sigma}_j)-\boldsymbol{\sigma}_i\cdot\boldsymbol{\sigma}_j$ and $h(r_{ij})=1+3/(m_\pi r_{ij})+3/(m_\pi r_{ij})^2$. 

To regularize  Eq.~(\ref{eq:mag_continuum}) on the lattice, we first analytically transform it to the momentum space, then discretize the momentum through the periodic boundary condition, and finally make fast Fourier transformations to obtain the lattice operator. The resulting expression for the $k$th component of the intrinsic term is, 
\begin{eqnarray}
\label{eq:intrinsic}
\mu_{\mathrm{intr}, k} = \mu_{N}\sum_{\boldsymbol{n}_1,\boldsymbol{n}_2} \epsilon_{I_1I_2 z}V_{\mathrm{intr},k}^{S_1S_2}(\boldsymbol{n}_1-\boldsymbol{n}_2):\rho_{S_1,I_1}(\boldsymbol{n}_1)\rho_{S_2,I_2}(\boldsymbol{n}_2):,
\end{eqnarray}
with
\begin{eqnarray} V_{\mathrm{intr},k}^{S_1S_2}(\boldsymbol{n}_1-\boldsymbol{n}_2)&=& \frac{m_N g_A^2}{4F_\pi^2}\sum_{\boldsymbol{p}}\frac{2\epsilon_{S_1 S_2 S_3} p_{S_3}p_k-\epsilon_{S_1S_2 k}(p^2-m_\pi^2)}{(p^2+m_\pi^2)^2}e^{i\boldsymbol{p}\cdot(\boldsymbol{n}_1-\boldsymbol{n}_2)}.
\end{eqnarray}
While for the Sachs term,
\begin{eqnarray}
\label{eq:Sachs}
\mu_{k,\mathrm{Sachs}} = \mu_N \sum_{\boldsymbol{n}_1,\boldsymbol{n}_2}\epsilon_{I_1I_2z}\epsilon_{ijk}\frac{n_{1,i}+n_{2,i}}{2}V_{\mathrm{Sachs},j}^{S_1 S_2}(\boldsymbol{n}_1-\boldsymbol{n}_2):\rho_{S_1I_1}(\boldsymbol{n}_1)\rho_{S_2I_2}(\boldsymbol{n}_2):,
\end{eqnarray}
with
\begin{eqnarray}  V_{\mathrm{Sachs},j}(\boldsymbol{n}_1-\boldsymbol{n}_2)=\frac{im_N g_A^2}{4F_\pi^2}\sum_{\boldsymbol{p}}
\left(\frac{\delta_{j,S1}p_{S_2}+\delta_{j,S_2}p_{S_1}}{p^2+m_\pi^2}-\frac{2p_{S_1}p_{S_2}p_j}{(p^2+m_\pi^2)^2}\right)e^{i\boldsymbol{p}\cdot(\boldsymbol{n}_1-\boldsymbol{n_2})}.
\end{eqnarray}
One can prove that in the continuum limit $a\rightarrow 0$ and $L\rightarrow \infty$, Eq.~(\ref{eq:intrinsic}) and~(\ref{eq:Sachs}) recover to Eq.~(\ref{eq:mag_decomp}) and~(\ref{eq:mag_continuum}).

\subsection{B.  The expression of perturbative contributions to the magnetic moment}

In this part, we  give the explicit expressions of the perturbative terms $\mu_{\mathrm{1N}}^{(0)},\mu_{\mathrm{1N}}^{(1)}$ and $\mathrm{\mu}_{2\mathrm{N}}^{(0)}$ in Eq.~(\ref{eq:PTmethod}). We introduce the transfer matrix of the non-perturbative Hamiltonian $H_S$,
\begin{eqnarray} \mathcal{M}_S \equiv :e^{-H_S a_t}:,
\end{eqnarray}
and define 
\begin{eqnarray}
&&|\Phi_{J,M=J}^{(0)}(L_t)\rangle\equiv \mathcal{M}_S^{L_t/2}|\Psi^T_{J,M=J}\rangle ,\nonumber\\
 &&|\Phi_{J,M=J}^{(1)}(L_t)\rangle\equiv\sum_{k=1}^{L_t/2}\mathcal{M}_S^{k-1}(\mathcal{M}-\mathcal{M}_S)\mathcal{M}_S^{L_t/2-k}|\Psi^T_{J,M=J}\rangle,\nonumber\\
\end{eqnarray}
then following the same derivation of \cite{PTmethod1}, one can find that the expressions of the perturbative terms as follows
\begin{eqnarray} 
\label{eq:PT_expression}&&\mu_{\mathrm{1N}/\mathrm{2N}}^{(0)}=\lim_{L_t \rightarrow \infty}\frac{\langle \Psi_{J,M=J}^{(0)}(L_t)|\mu_{\mathrm{1N}/\mathrm{2N},z}|\Psi_{J,M=J}^{(0)}(L_t)\rangle}{\langle \Psi_{J,M=J}^{(0)}(L_t)|\Psi_{J,M=J}^{(0)}(L_t)\rangle},\nonumber\\
&&\mu_{\mathrm{1N}}^{(1)}=\lim_{L_t \rightarrow \infty} \left(\frac{\langle \Psi_{J,M=J}^{(0)}(L_t)|\mu_{\mathrm{1N},z}|\Psi_{J,M=J}^{(1)}(L_t)\rangle+\mathrm{c.c.}}{\langle \Psi_{J,M=J}^{(0)}(L_t)|\Psi_{J,M=J}^{(0)}(L_t)\rangle}-\mu_{\mathrm{1N}}^{(0)}\frac{\langle \Psi_{J,M=J}^{(0)}(L_t)|\Psi_{J,M=J}^{(1)}(L_t)\rangle+\mathrm{c.c.}}{\langle \Psi_{J,M=J}^{(0)}(L_t)|\Psi_{J,M=J}^{(0)}(L_t)\rangle}\right)
\end{eqnarray}
The expressions above are calculated through the auxiliary field quantum Monte Carlo method, which makes use of auxiliary field transformation to decouple the complex many-body correlations into the fluctuation of auxiliary fields, and we refer interested readers to Ref.~\cite{Wang:2026} for details.

\subsection{C. Trial state preparation}

As discussed in the main text, we use multi-reference trial states composed of different shell-model Slater determinants and having the definite spin $J$ in the continuum limit, see Eq.~(\ref{eq:TrialState}). Compared to a single Slater determinant frequently used as the trial state in previous NLEFT calculations, our construction of trial state significantly accelerates the convergence of lattice correlation functions versus the projection time $\tau$~\cite{Wang:2026}. The reason is that the single Slater determinant usually mixes excited states with spin $J'\neq J$,  which may give large excited-state contamination when calculating non-scalar observables such as the magnetic moment. More specifically, we take the calculation of $^8$Li as an example. For the ground state of  $^8$Li with $J^\pi=2^+$, its multi-reference trial state $|\Psi^T_{J=M=2}\rangle$ consists of two Slater determinants $|\Psi_{M=2}^1\rangle$ and  $|\Psi_{M=2}^2\rangle$ shown in Figure~\ref{fig:trial_state_Li8}. It is easy to see that both $|\Psi_{M=2}^1\rangle$ and  $|\Psi_{M=2}^2\rangle$ has a mixture of $J=3$ component.  Therefore, if one uses either of them as the trial state to compute the magnetic moment, the rank-one tensor operator $\mu_z$ mixes the ground state with the $J=3$ excited states. Consequently, the asymptotic behavior of the ratio on the R.H.S. of Eq.~(\ref{eq:Master}) is $\mu+ce^{-\Delta E\tau/2}$, with $c$  the non-zero coefficient originated from this mixing and $\Delta E$  the energy gap between the ground state and the $3^+_1$ state of $^8$Li. Since experimental value $\Delta E=2.26$ MeV~\cite{Tilley:2004} is small, the  exponential $e^{-\Delta E\tau/2}$ decays very slowly, resulting in large excited-state effects. On the other hand, through a proper combination of $|\Psi_{M=2}^1\rangle$ and  $|\Psi_{M=2}^2\rangle$, the contamination from $J=3$ excited states is removed from the multi-reference trial state $|\Psi_{J=M=2}^{\mathrm{T}}\rangle$   and the signal converges much more rapidly.  For illustration, we calculate $\mu_{\mathrm{1N}}^{(0)}$ of $^8$Li using $|\Psi^T_{J=M=2}\rangle$,  $|\Psi_{M=2}^1\rangle$ and $|\Psi_{M=2}^2\rangle$, separately, and show their respective evolution versus $\tau$  in Figure~\ref{fig:compare_trialstate_Mag}. For the multi-reference trial state  $|\Psi^T_{J=M=2}\rangle$, the result reaches the  plateau at around $\tau=0.4$~MeV$^{-1}$. In comparison, the data point of the two single Slater determinants is not fully converged even at $\tau=0.6$~MeV$^{-1}$, due to the remaining excited-state mixing. This would induce significantly enhanced extrapolation uncertainties.

\begin{figure}[htb]
\centering
\includegraphics[width=0.7\textwidth,angle=0]{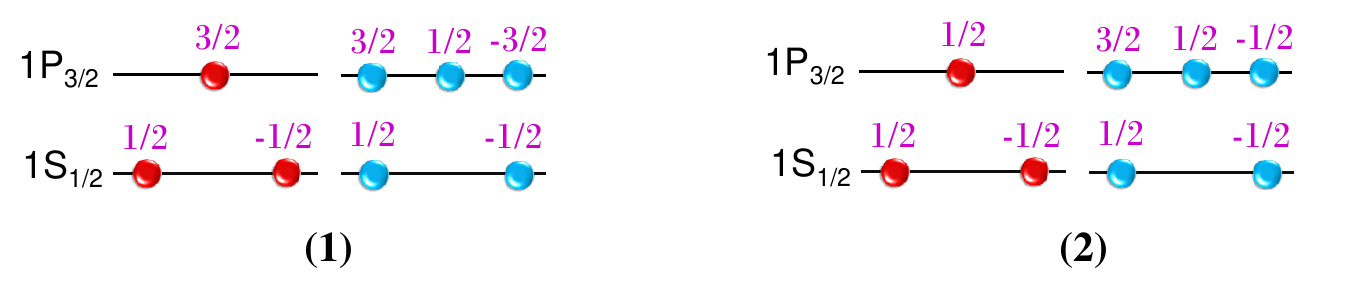}
\caption{The two shell-model Slater determinants, $|\Psi_{M=2}^1\rangle$ (left panel) and $|\Psi_{M=2}^2\rangle$ (right panel), that form the multi-reference trial state $|\Psi^T_{J=M=2}\rangle$ of $^{8}$Li used in this work.  The red (blue)
circle represents the occupied orbit of the proton (neutron).
The value close to the circle denotes the magnetic quantum
number of the orbit.  }
\label{fig:trial_state_Li8}
\end{figure}

\begin{figure}[htb]
\centering
\includegraphics[width=0.7\textwidth,angle=0]{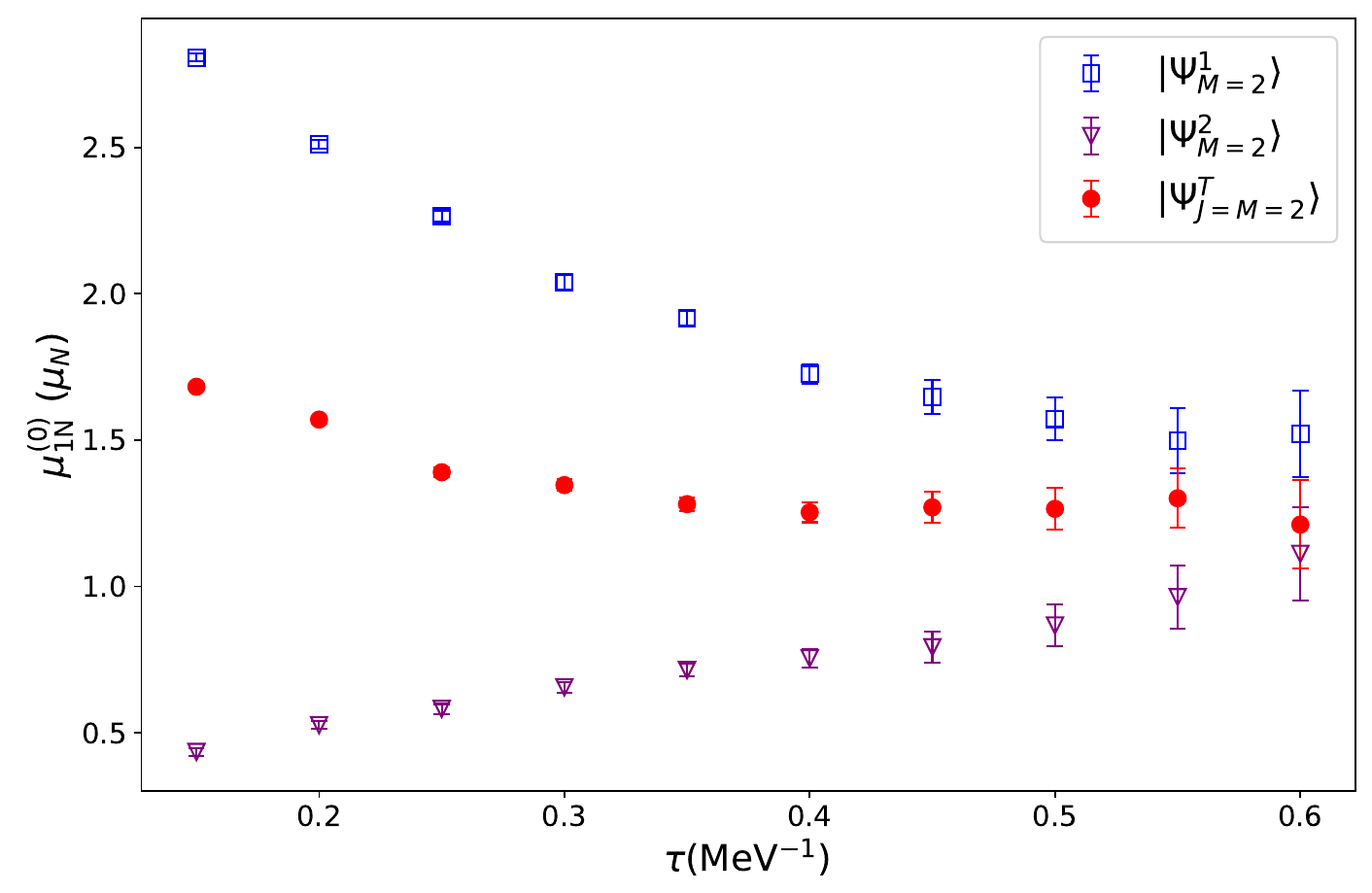}
\caption{The evolution of $\mu_{1\mathrm{N}}^{(0)}$ of $^8$Li ground state versus the projection time $\tau$, calculated from the three trial states $|\Psi^T_{J=M=2}\rangle$,  $|\Psi_{M=2}^1\rangle$ and $|\Psi_{M=2}^2\rangle$. The error bar represents statistical uncertainty.}
\label{fig:compare_trialstate_Mag}
\end{figure}

\subsection{D. Imaginary time extrapolation }

For the magnetic moment $\mu$, we fit to its perturbative components $\mu_{\mathrm{1N}}^{(0)}$, $\mu_{\mathrm{1N}}^{(1)}$ and $\mu_{\mathrm{2N}}^{(0)}$ separately. For $\mu_{\mathrm{1N}}^{(0)}$ and $\mu_{\mathrm{2N}}^{(0)}$,  we fit the calculated values at finite projection time $\tau$ using either a constant or a single-exponential ansatz,
\begin{equation}
    \mu(\tau) = \mu+ce^{-d\tau/2},
\end{equation}
depending on the convergence of the data in the asymptotic region. For $\mu_{\mathrm{1N}}^{(1)}$, due to the rapid growth of statistical error, $\tau$ cannot be taken too large and even a single-exponential ansatz would be too complicated for the fit. Therefore, we stick to making a constant fit to extract $\mu_{\mathrm{1N}}^{(1)}$. For illustration, we show the lattice data and the fit of $^{27}$Al in Figure~\ref{fig:Al27_fit}.

\begin{figure}[htb]
\centering
\includegraphics[width=1.0\textwidth,angle=0]{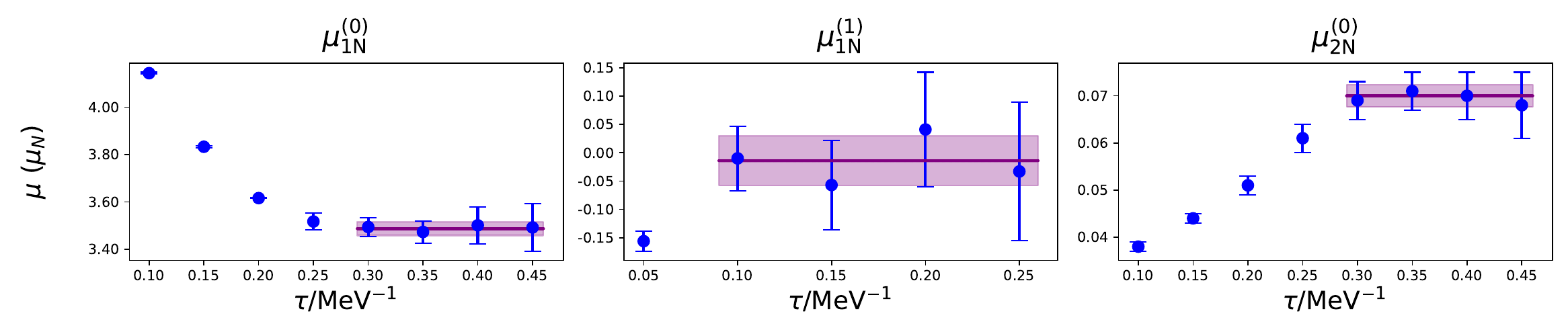}
\caption{Imaginary time extrapolation for the magnetic moment of $^{27}$Al. The three columns represent $\mu_{\mathrm{1N}}^{(0)}$, $\mu_{\mathrm{1N}}^{(1)}$ and $\mu_{\mathrm{2N}}^{(0)}$ respectively. The blue points are lattice data  with error bars the statistical uncertainty. The purple shaded region denotes the  result of extrapolation and the range of data used for fit.}
\label{fig:Al27_fit}
\end{figure}

\subsection{E. Analysis of rotational symmetry breaking}

In this section, we estimate the influence of  rotational symmetry breaking on the lattice calculation of the magnetic moment, using one representative nucleus for each shell, i.e., $^{3}$H of $s$-shell, $^9$C of $p$-shell and $^{27}$Al of $sd$-shell.

We first analyze the effect induced by the finite size of the lattice. We calculate  $\mu_{\mathrm{1N}}^{(0)}$ for different values of the box length $L$ at the same projection time $\tau=0.2$ MeV$^{-1}$, using the same interaction $H_S$. The result is presented in Fig.~\ref{fig:FV_effect}. For $^{3}$H, the finite-volume effect is already negligible at $L=7a$. For $^{9}$C, the finite-volume effect appears at $L=7a$ but cannot be distinguished from statistical errors at $L=9a$.  For $^{27}$Al, the artifact is significant for $L<9a$ but becomes small for $L=9a$. We also checked other nuclei, and find that the finite volume effect on magnetic moment tends to be enhanced as the mass number increases, but is generally under good control for $L\ge 9a\approx 12~\mathrm{fm}$ used in this work.

\begin{figure}[htb]
\centering
\includegraphics[width=1.0\textwidth,angle=0]{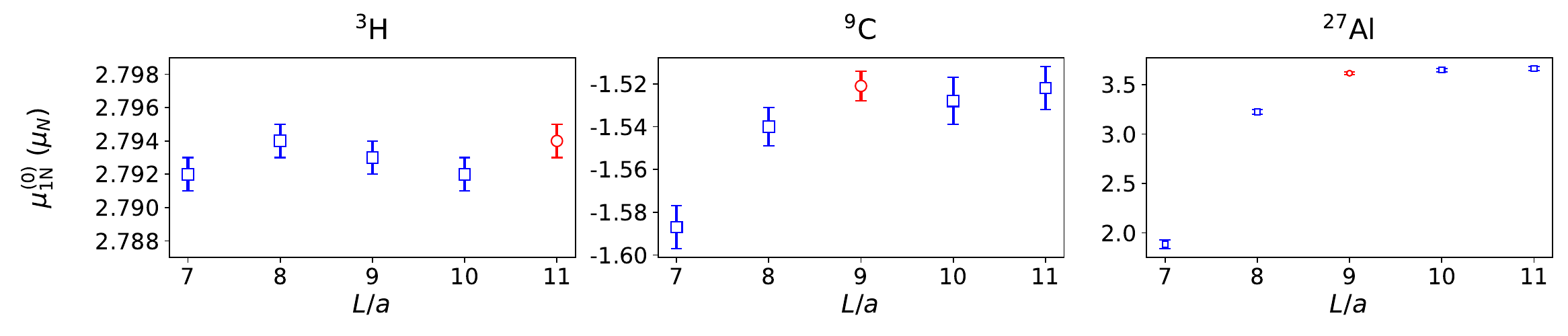}
\caption{The dependence of $\mu_{1\mathrm{N}}^{(0)}$ on the box length $L$ for $^{3}$H, $^{9}$C and $^{27}$Al. All the data points are calculated at the same projection time $\tau=0.2$~MeV$^{-1}$, using the same interaction $H_S$. The red point denotes the data used in the main text. The error bars denote statistical uncertainty. }
\label{fig:FV_effect}
\end{figure}

Another origin for the breaking of rotational symmetry on the lattice is the finite lattice spacing, and the most straightforward way to reduce this error, in principle, is to extrapolate to the limit $a=0$. However, this approach is complicated by the non-renormalizable nature of $\chi$EFT where uncancelled ultraviolet divergences remain~\cite{NLEFT_rotationsymmetry1}. In Ref.~\cite{NLEFT_rotationsymmetry2}, an alternative method is proposed to quantify the impact of non-zero lattice spacing on the matrix element of irreducible tensor operators sandwiched by a pair of
bound state wave functions, such as the magnetic moment studied here. In this approach, one considers the following irreducible matrix element of $\mu_z$, 
\begin{equation} 
\label{eq:xxx}(\Psi_{JM}||\mu_z||\Psi_{JM})=\frac{\langle\Psi_{JM}|\mu_z|\Psi_{JM}\rangle}{\langle JM|J1;M0\rangle},
\end{equation}
where $|\Psi_{JM}\rangle$ is the nuclear ground state calculated through NLEFT and carries the quantum numbers $J$ and $M$ in the continuum limit. $\langle JM|J1;M0\rangle$ is the corresponding Clebsch-Gordan coefficient. Following Wigner-Eckart theorem, $(\Psi_{JM}||\mu_z||\Psi_{JM})$ is independent of $M$ as $a\rightarrow 0$ and $L\rightarrow \infty$. For $a \neq 0$,  $(\Psi_{JM}||\mu_z||\Psi_{JM})$ depends on $M$ and the splittings between different components of Eq.~(\ref{eq:xxx}). Therefore, it serves as a sensitive indicator of rotational
symmetry breaking effects. In Table~\ref{tab:RSB_quatify}, we compare the values of $(\Psi_{J,M=J}||\mu_z||\Psi_{J,M=J})$ and $(\Psi_{J,M=J-1}||\mu_z||\Psi_{J,M=J-1})$ for the above three nuclei, calculated using the LO Hamiltonian $H_S$. It can be seen that for $^{3}$H and $^{9}$C, the difference between the two components is  almost zero compared to the statistical uncertainty, indicating negligible rotational symmetry breaking effects. For $^{27}$Al, the difference is larger but still at a few percent level compared to the central value.

\begin{table}
\begin{center}
\renewcommand\arraystretch{1.4}
\resizebox{300pt}{!}{
		\begin{tabular}{|c|c|c|c| }
			\hline           \diagbox{$(\Psi_{JM}||\mu_z||\Psi_{JM})$}{nucleus($J^\pi$)}&$^{3}$H(1/2$^+$)&$^{9}$C(3/2$^-$)&$^{27}$Al(5/2$^+$)\\
        \hline
        $M=J$&4.839(2)&-1.937(13)&4.339(17)\\
        \hline
        $M=J-1$&4.839(2)&-1.942(27)&4.209(28)\\
        \hline
		\end{tabular}}
  \end{center}
\caption{The values of $(\Psi_{J,M=J}||\mu_z||\Psi_{J,M=J})$ and $(\Psi_{J,M=J-1}||\mu_z||\Psi_{J,M=J-1})$ for $^3$H, $^9$C and $^{27}$Al, calculated through the LO Hamiltonian $H_S$ and at the same projection time $\tau=0.2\mathrm{MeV}^{-1}$ without extrapolation.}
	\label{tab:RSB_quatify}
\end{table}

\begin{table}
\begin{center}
\renewcommand\arraystretch{1.6}
\resizebox{300pt}{!}{
		\begin{tabular}{|c|c|c|c|c|c| }
			\hline
			      Nucleus& $\mu_{1\mathrm{N}}^{(0)}$ &$\mu_{1\mathrm{N}}^{(1)}$ &$\mu_{2\mathrm{N}}^{(1)}$ &$\mu_{\mathrm{latt}}$ &$\mu_{\mathrm{exp}}$   \\
                 \hline
                 $^{2}$H(1$^+$)&0.874(2)&-0.008(6)&0&0.866(6)&0.857\\
		\hline
        $^{3}$H(1/2$^+$)&2.791(2)&0.021(10)&0.139(1)&2.951(10)&2.979\\
		\hline     $^{3}$He(1/2$^+$)&-1.936(3)&-0.016(10)&-0.138(1)&-2.090(10)&-2.128\\
        \hline
        $^{6}$Li(1$^+$)&0.843(18)&0.044(19)&0&0.887(26)&0.822\\
        \hline
        $^{7}$Li(3/2$^-$)&3.146(12)&-0.019(16)&0.121(1)&3.248(20)&3.256\\
        \hline
        $^{8}$Li(2$^+$)&1.252(21)&0.317(38)&0.115(2)&1.684(43)&1.654\\
        \hline
        $^{9}$Li(3/2$^-$)&3.397(9)&-0.059(9)&0.130(2)&3.468(13)&3.437\\
        \hline
        $^{11}$Li(3/2$^-$)&3.656(9)&-0.038(20)&0.149(2)&3.767(22)&3.671\\
        \hline
        $^{7}$Be(3/2$^-$)&-1.364(8)&-0.025(10)&-0.121(1)&-1.461(22)&-1.399\\
        \hline
        $^{9}$Be(3/2$^-$)&-1.450(14)&0.212(35)&-0.002(1)&-1.240(38)&-1.177\\
        \hline
        $^{11}$Be(3/2$^-$)&-1.847(11)&0.158(18)&-0.011(1)&-1.700(21)&-1.682\\
        \hline
        $^{8}$B(2$^+$)&1.400(30)&-0.230(52)&-0.123(2)&1.047(60)&1.036\\
        \hline
        $^{10}$B(3$^+$)&1.823(4)&-0.003(15)&0&1.820(16)&1.800\\
        \hline
        $^{11}$B(3/2$^-$)&3.125(18)&-0.497(66)&0.105(2)&2.733(68)&2.688\\
        \hline
        $^{12}$B(1$^+$)&1.098(26)&-0.209(79)&0.060(2)&0.949(83)&1.003\\
        \hline
        $^{9}$C(3/2$^-$)&-1.518(9)&0.130(50)&-0.108(1)&-1.496(51)&-1.391\\
        \hline
        $^{11}$C(3/2$^-$)&-1.231(14)&0.424(78)&-0.097(2)&-0.904(81)&-0.964(1)\\
        \hline
		\end{tabular}}
  \end{center}
\caption{The extrapolation results of $\mu_{\mathrm{1N}}^{(0)},\mu_{\mathrm{1N}}^{(1)}, \mu_{\mathrm{2N}}^{(0)}$ and their sum $\mu_{\mathrm{latt}}$ for selected $A\le 12$ nuclei, compared to the experimental value $\mu_{\mathrm{exp}}$~\cite{Purcell:2010,Tilley:2002,Tilley:2004,Borremans:2005,Neugart:2008,Williams:1970, Millman:1939,Wolber:1970}. The number in the bracket denotes statistical error. All values are in unit of $\mu_N$. Since the two-body magnetic moment operator used in this work is an isospin vector, its contribution to isospin-scalar nuclei like $^{6}$Li and $^{10}$B vanishes.  }
	\label{tab:results_display_light}
\end{table}

\subsection{F. Table of numerical values}

\subsubsection{Magnetic moments}

In Table~\ref{tab:results_display_light} and~\ref{tab:results_display_Al}, we list  the numerical values of magnetic moments for selected light nuclei and aluminum isotopes, respectively. For comparison, we also list the experimental value. 

\subsubsection{Comparison with other $ab\ initio$ methods for light nuclei}

In Table~\ref{tab:comparison_between_methods}, we give the values of the calculated magnetic moments for selected light nuclei using different $ab\ initio$ methods, corresponding to  Figure~\ref{fig:Methods_compare_spshell} in the main text. We also list the experimental values for comparison.

\begin{table}
\begin{center}
\renewcommand\arraystretch{1.6}
\resizebox{400pt}{!}{
		\begin{tabular}{|c|c|c|c|c|c| }
			\hline
			      Nucleus& $\mu_{1\mathrm{N}}^{(0)}$ &$\mu_{1\mathrm{N}}^{(1)}$ &$\mu_{2\mathrm{N}}^{(1)}$ &$\mu_{\mathrm{latt}}$ &$\mu_{\mathrm{exp}}$   \\
                 \hline
        $^{23}$Al(5/2$^+$)&4.021(28)$_{\mathrm{sta}}$&-0.163(40)$_{\mathrm{sta}}$&0.014(1)$_{\mathrm{sta}}$&3.872(49)$_{\mathrm{sta}}$(77)$_{\mathrm{rot}}$(92)$_{\mathrm{tot}}$&3.89(2)\\
        \hline
        $^{24}$Al(1$^+$)&2.401(156)$_{\mathrm{sta}}$&0.339(98)$_{\mathrm{sta}}$&0.067(4)$_{\mathrm{sta}}$&2.807(184)$_{\mathrm{sta}}$(56)$_{\mathrm{rot}}$(193)$_{\mathrm{tot}}$&2.99(9)\\
        \hline
        $^{25}$Al(5/2$^+$)&3.804(9)$_{\mathrm{sta}}$&-0.113(48)$_{\mathrm{sta}}$&0.019(1)$_{\mathrm{sta}}$&3.710(49)$_{\mathrm{sta}}$(74)$_{\mathrm{rot}}$(89)$_{\mathrm{tot}}$&3.645(1)\\
        \hline
        $^{26}$Al(5$^+$)&2.797(8)$_{\mathrm{sta}}$&0.009(9)$_{\mathrm{sta}}$&0&2.806(12)$_{\mathrm{sta}}$(56)$_{\mathrm{rot}}$(57)$_{\mathrm{tot}}$&2.803(4)\\
        \hline
        $^{27}$Al(5/2$^+$)&3.487(29)$_{\mathrm{sta}}$&-0.014(44)$_{\mathrm{sta}}$&0.070(2)$_{\mathrm{sta}}$&3.543(50)$_{\mathrm{sta}}$(71)$_{\mathrm{rot}}$(86)$_{\mathrm{tot}}$&3.641\\
        \hline
        $^{28}$Al(3$^+$)&3.590(33)$_{\mathrm{sta}}$&-0.398(64)$_{\mathrm{sta}}$&0.106(3)$_{\mathrm{sta}}$&3.298(72)$_{\mathrm{sta}}$(66)$_{\mathrm{rot}}$(98)$_{\mathrm{tot}}$&3.241(5)\\
        \hline
        $^{29}$Al(5/2$^+$)&3.655(30)$_{\mathrm{sta}}$&-0.266(37)$_{\mathrm{sta}}$&0.094(1)$_{\mathrm{sta}}$&3.483(48)$_{\mathrm{sta}}$(70)$_{\mathrm{rot}}$(84)$_{\mathrm{tot}}$&3.665(2)\\
        \hline
        $^{30}$Al(3$^+$)&3.352(34)$_{\mathrm{sta}}$&-0.488(62)$_{\mathrm{sta}}$&0.042(4)$_{\mathrm{sta}}$&2.906(71)$_{\mathrm{sta}}$(58)$_{\mathrm{rot}}$(92)$_{\mathrm{tot}}$&3.012(7)\\
        \hline
        $^{31}$Al(5/2$^+$)&3.851(24)$_{\mathrm{sta}}$&-0.062(36)$_{\mathrm{sta}}$&0.108(1)$_{\mathrm{sta}}$&3.897(43)$_{\mathrm{sta}}$(78)$_{\mathrm{rot}}$(89)$_{\mathrm{tot}}$&3.832(5)\\
        \hline
		\end{tabular}}
  \end{center}
\caption{The extrapolation results of $\mu_{\mathrm{1N}}^{(0)},\mu_{\mathrm{1N}}^{(1)}, \mu_{\mathrm{2N}}^{(0)}$ and their sum $\mu_{\mathrm{latt}}$ for Al isotopes with $23 \le A\le 31$, compared to the experimental value $\mu_{\mathrm{exp}}$~\cite{Ozawa:2006,Minamisono:1976,Cooper:1996,Nishimura:2007,Al33}. The numbers in the bracket with subscripts `sta', `rot' and `tot' denote  the statistical uncertainty,  rotational-symmetry-breaking uncertainty and full uncertainty, respectively. All values are in units of $\mu_N$. Since the two-body magnetic moment operator used in this work is an isospin vector, its contribution to the isospin-scalar nucleus $^{26}$Al vanishes. }
	\label{tab:results_display_Al}
\end{table}

\begin{table}
\begin{center}
\renewcommand\arraystretch{1.6}
\resizebox{300pt}{!}{
		\begin{tabular}{|c|c|c|c|c|c| }
			\hline
			      \diagbox{Nucleus}{Method}& VMC &NCSM &GFMC &NLEFT &EXP   \\
                 \hline
                 $^{2}$H(1$^+$)&---&---&---&0.866(6)&0.857\\
		\hline
        $^{3}$H(1/2$^+$)&2.970&---&2.924&2.951(10)&2.979\\
		\hline     $^{3}$He(1/2$^+$)&-2.116&---&-2.089&-2.090(10)&-2.128\\
        \hline
        $^{6}$Li(1$^+$)&0.840&0.843(2)&0.937&0.887(26)&0.822\\
        \hline
        $^{7}$Li(3/2$^-$)&3.301&3.02(2)&3.307&3.248(20)&3.256\\
        \hline
        $^{8}$Li(2$^+$)&1.660&1.42(4)&1.55(8)&1.684(43)&1.654\\
        \hline
        $^{9}$Li(3/2$^-$)&3.494&2.98(5)&3.29(16)&3.468(13)&3.437\\
        \hline
        $^{11}$Li(3/2$^-$)&---&3.54(4)&---&3.767(22)&3.671\\
        \hline
        $^{7}$Be(3/2$^-$)&-1.443&-1.15(1)&-1.37(2)&-1.461(22)&-1.399\\
        \hline
        $^{9}$Be(3/2$^-$)&-1.06(6)&0.212(35)&-1.19(3)&-1.240(38)&-1.177\\
        \hline
        $^{11}$Be(3/2$^-$)&---&-1.47(3)&---&-1.700(21)&-1.682\\
        \hline
        $^{8}$B(2$^+$)&1.028&---&1.09(11)&1.047(60)&1.036\\
        \hline
        $^{10}$B(3$^+$)&1.839&1.836&---&1.820(16)&1.800\\
        \hline
        $^{11}$B(3/2$^-$)&---&2.371&---&2.733(68)&2.688\\
        \hline
        $^{12}$B(1$^+$)&---&0.561&---&0.949(83)&1.003\\
        \hline
        $^{9}$C(3/2$^-$)&-1.590&---&-1.38(20)&-1.496(51)&-1.391\\
        \hline
        $^{11}$C(3/2$^-$)&---&-0.638&---&-0.904(81)&-0.964(1)\\
        \hline
		\end{tabular}}
  \end{center}
\caption{The list of data for the  magnetic moments of selected light nuclei calculated through VMC~\cite{ChambersWallPRL, ChambersWallPRC}, NCSM~\cite{Forssen:2009,NCSM_B,NCSM_C}, GFMC~\cite{ChambersWallPRL, ChambersWallPRC} and NLEFT, and their comparison to the experimental value. The theoretical uncertainty is given in the bracket. All values are in units of $\mu_N$. }
	\label{tab:comparison_between_methods}
\end{table}

\end{widetext}
\end{document}